\documentclass[prl,twocolumn]{revtex4}
\usepackage{graphicx}
\begin{document}
\newcommand{\x}{{\bf r}}
\newcommand{\K}{{\bf k}}

\title{Single quantum realization of a collision of two Bose--Einstein condensates}

\author{Jan Chwede\'{n}czuk$^1$,  Pawe{\l} Zi\'{n}$^1$, Kazimierz
Rz\c{a}\.zewski$^3$ and Marek Trippenbach$^{1,2}$.}

\affiliation{$^1$Institute for Theoretical Physics, Warsaw
University, Ho\.{z}a 69, PL-00-681 Warsaw, Poland, \\ $^{2}$Soltan
Institute for Nuclear Studies, Ho\.{z}a 69, PL-00-681 Warsaw,
Poland\\ $^3$Center for Theoretical Physics, Polish Academy of
Sciences, Al. Lotnik\'{o}w 32/46, PL-02-668 Warsaw, Poland.}

\begin{abstract}
We propose a method for simulating a single realization of a
collision of two Bose--Einstein condensates. Recently in Zi\'{n}
{\it et al.} (Phys. Rev. Lett. {\bf 94}, 200401 (2005)) we
introduced a quantum model of an incoherent elastic scattering in a
collision of two counter-propagating atomic Gaussian wavepackets.
Here show that this model is capable of generating the data that can
be interpreted as results of a single collision event. We find a
range of parameters, including relative velocity, population and the
size of colliding condensates, where the structure of hallo of
scattered atoms in a single realization strongly differs from that
averaged over many realizations.
\end{abstract}

\maketitle

According to the standard interpretation of quantum mechanics the
single particle wavefunction $\psi({\bf r},t)$ describes probability
amplitude and $|\psi({\bf r},t)|^2$ - probability density for
finding this particle around $\bf r$ at time $t$. To make a
comparison between theory and experiment one needs to repeat the
same measurement many times since the agreement can be checked only
on the statistical level. Recent developments in the physics of Bose
Einstein condensates of neutral atom gasses \cite{Dalfovo} created
an extremely interesting situation where already a single
measurement, due to large number of particles involved, reveals some
statistical aspects. One prominent example is given by two
overlapping independent Bose-Einstein condensates (each of them in
the Fock state), where every single measurement of the system
reveals an interference pattern. At the same time however, no
pattern is present in the average over many measurements
\cite{prazki,java}. This example shows that even for many particle
systems there is a substantial difference between results of a
single experiment and the statistical average (over many repetitions
of the experiment). The latter, in analogy with the single particle
case, is much easier to handle theoretically, the former, more
straightforward for experimentalist, is a real challenge for
theoretician. Nevertheless, in some cases theoretical solutions are
possible. Javanainen and Yoo \cite{java} were able to prove that
indeed interference fringes are present in single measurement of
overlapping independent condensates. Bach and Rz\c{a}\.zewski
\cite{Bach} studied quantum correlations in atomic systems, and
noted that single measurements data contain information about these
correlations. Dziarmaga \cite{Sacha} proved that any Bogoliubov
vacuum can be brought to a diagonal form in a time-dependent
orthonormal basis. This diagonal form is tailored for simulations of
quantum measurements on an excited condensate. They illustrated
their method using example of the phase imprinting of a dark
soliton. Here we examine a similar approach for the case when the
substantial depletion of the condensate occurs.

Bragg diffraction divides a Bose-Einstein condensate into two
initially overlapping components, which immediately start moving
away from each other with high relative momentum. Elastic collisions
between pairs of atoms from distinct wave packets cause losses that
can significantly deplete the condensate. Recently \cite{Zin,Zin2}
we introduced the quantum model of two counter-propagating atomic
Gaussian wave packets to study such collisions and the transition
from spontaneous to stimulated regime. Within this model correlation
functions of all orders are accessible. It is then not surprising
that it allows for a next conceptual step - reconstruction of the
result in a single realization of the experiment. The proposed
method is valid in the regime of bosonic stimulation and we show the
relation to the previous analytical methods \cite{Sacha}, but also
how it can be used to test the validity of stochastic method
\cite{Norie}.

We consider a half-collision of two Bose-Einstein condensates
initiated by Bragg scattering. Hamiltonian of the system in second
quantization reads, see \cite{Zin,Zin2}
\begin{eqnarray}
\label{ham1}\mathrm{\hat H}&=&-\int d^3 r \, \hat \Psi^\dagger({\bf
r},t)\frac{\hbar^2\nabla^2}{2m}\hat\Psi({\bf r},t)\nonumber\\
&+&\frac{g}{2}\int \mbox{d}^3 r \, \hat{\Psi}^\dagger({\bf r},t)
\hat{\Psi}^ \dagger({\bf r},t)\hat{\Psi}({\bf r},t) \hat{\Psi}({\bf
r},t),
\end{eqnarray}
where $\hat\Psi(\x,t)$ is the bosonic field operator. Atoms interact
via two-body contact interaction, determined by the single coupling
constant $g$. As the system consists of two counter-propagating,
highly occupied atomic wave-packets and the ``sea'' of unoccupied
modes, in the spirit of Bogoliubov approximation we decompose the
field operator into $c-$number condensate wave-function $\psi_Q(\x,
t)+\psi_{-Q}(\x, t)$ and the quantum field of scattered atoms,
$\hat\delta(\x,t)$, :
\begin{eqnarray}
\hat\Psi(\x, t)&=&\psi_Q(\x, t)+\psi_{-Q}(\x, t)+\hat{\delta}(\x, t).
\label{deco1}
\end{eqnarray}
The subscripts $\pm Q$ denote the mean momentum of the colliding
wavepackets. In what follows we assume that the condensate
wave-function satisfies independently the time-dependent GPE.  We
insert Eq.~(\ref{deco1}) into the Hamiltonian (\ref{ham1}) and keep
only terms, that lead to creation or annihilation of a pair of
particles in the $\hat\delta$ field of scattered atoms,
\begin{eqnarray}
\label{ham2} \mathrm{H}&=&-\int \mbox{d}^3 r \, \hat{\delta}^
\dagger({\bf r},t)\frac{\hbar^2\nabla^2}{2m}\hat{\delta}({\bf r},t)
\\ \nonumber
&+&g\int \mbox{d}^3 r \, \hat{\delta}^\dagger({\bf r},t)
\hat{\delta}^\dagger({\bf r},t) \psi_Q({\bf r},t) \psi_{-Q}({\bf r},t)+
\mathrm{H.c.}
\end{eqnarray}
Further simplification is possible in the regime when the
collisional time $t_C=\sigma/(\hbar Q/m)$ is much shorter than the
linear disspersion time $t_{LD}=m\sigma^2/\hbar$ and the nonlinear
disspersion time $t_{ND}=\sqrt{\pi^{3/2} m\sigma^5/gN}$ \cite{Zin2}.
Here $\sigma$ and $N/2$ is the radius and number of particles in
each of the colliding partners. In this regime neither dispersion
nor the nonlinearity produce any noticeable effects on the
condensate part during the collision and the change of the
condensates' shape can be neglected. To model the condensates we use
spherically symmetric Gaussian wave-functions
\begin{eqnarray}
\label{cond} &&\psi_{\pm Q}({\bf r},t) = \sqrt{\frac{N}{2\pi^{3/2}\sigma^{3}}}
\exp \left[ \pm
iQx_1
-\frac{i\hbar t Q^2}{2m} \right] \times\nonumber\\
&&\times\exp\left[-\frac{1}{2\sigma^2}\left(\left(x_1\mp\frac{\hbar
Qt}{m}\right)^2+x_2^2+x_3^2\right)\right],
\end{eqnarray}
where ${\bf r}=(x_1,x_2,x_3)$. In this approximation the Heisenberg equation for the
dimensionless field operator $\hat{\delta}({\bf r},t)\,\sigma^{3/2}\exp \left(i \beta
t/2\right) \rightarrow \hat{\delta}({\bf r},t) $ in the dimensionless units
($t/t_C\rightarrow t$, $x_i/\sigma \rightarrow  x_i$ (for $i=1,2,3$), and with parameters
$t_{LD}/t_C=\beta$ and $(t_{LD}/t_{ND})^2=\alpha$)  reads
\begin{equation}
\label{evo}i\beta\partial_t\hat{\delta}(\x,t)=-\frac{1}{2}\left(\nabla^2+\beta^2\right)
\hat{\delta}(\x,t)
+\alpha e^{-r^2-t^2}\hat{\delta}^\dagger(\x,t).\label{EofM}
\end{equation}
We decompose the field operator into a basis modes of spherically symmetric harmonic
oscillator \cite{Abram},
\begin{eqnarray}
\label{decomp}\hat{\delta}(\x,t)&=&\sum_{n,l,m}R_{n,l}(r)Y_{lm}
(\theta,\phi)\hat b_{n,l,m}(t),
\end{eqnarray}
as discussed in \cite{Zin}, where we solved the system of linear
equations for operators $b_{n,l,m}(t)$.

Here, following the ideas introduced in \cite{Braunstein}, we take
the advantage of the fact that the Hamiltonian of the system is
quadratic in creation and anihilation operators and reduce the
problem to independently evolving modes that experience squeezing.
To do so we map all three quantum numbers $n,l,m$ into single index
($[n,l,m] \rightarrow [i]$) and write the solution of the evolution
equation (\ref{evo}) in a compact form
\begin{eqnarray}
\label{vectors}\vec b(t)=C\vec b(0)+S\vec b^\dagger(0),
\end{eqnarray}
where vector $\vec b=(\hat b_1,\hat b_2,\ldots)^T$ and $C$ and $S$
are time dependent (evolution) matrices.
According to \cite{Braunstein}, $C$ and $S$ can be written as
$C=UC_DV^\dagger$ and $S=US_DV^\mathrm{T}$, were $U$ and $V$ are
unitary and $C_D$ and $S_D$ are diagonal with non-negative
eigenvalues.
Hence it is possible to define new initial modes (vectors)
$\vec{a}(0)=V^\dagger\vec b(0)$ and new final modes: $\vec{
a}(t)=U^\dagger\vec b(t)$ that are related by simple diagonal
transformation
\begin{equation}
\vec{a}(t)=C_D\vec{a}(0)+S_D\vec{a}^\dagger(0)\label{diagol}
\end{equation}
with a condition $C^2_{D,ii}-S^2_{D,ii}\equiv c^2_i-s^2_i=1$. The
choice of the time $t$ is crucial as it determines both initial and
final sets of modes in Eq.~(\ref{diagol}). It follows from
Eq.~(\ref{diagol}) that the system is in the multimode squeezed
state~\cite{Braunstein}, with atomic field operator
\begin{equation}\label{decompose}
\hat\delta(\x,t)=\sum_i\psi_i(\x,t)\hat a_i(t),
\end{equation}
where $\hat a_i(t)=c_i\hat a_i(0)+s_i\hat a^\dagger_i(0)$. Notice
that the density matrix is these modes diagonal~\cite{remark1}
\begin{eqnarray*}
\rho(\x_1,\x_2,t)=
\langle\hat\delta^\dagger(\x_1,t)\hat\delta(\x_2,t)\rangle=
\sum_i\psi_i^*(\x_1,t)\psi^{ }_i(\x_2,t)s_i^2.
\end{eqnarray*}
To find the set of $\psi_i(\x,t)$ we first solve the linear equation
(\ref{evo}), obtain the general solution of the form (\ref{vectors})
and evaluate the density matrix. Finally, following
reference~\cite{Sacha}, we find the squeezed modes by
diagonalization of the density matrix and the anomalous matrix.

For highly populated modes ($s_i\gg1$) we have $c_i\simeq s_i$. Then
the anihilation operators can be rewritten as
\begin{equation}
\hat a_i(t)\simeq s_i\left(\hat a^{ }_i(0)+\hat
a^\dagger_i(0)\right)=\sqrt{2\langle n_i\rangle}  \hat x_i(0),
\end{equation}
where the hermitian operator $\hat x_i=(\hat a^{ }_i+\hat
a_i^\dagger)/\sqrt2$  is the field quadraure and $\langle
n_i\rangle=s_i^2$ is the mode occupation. If we assume that highly
populated modes have dominant contribution to the field operator,
then
\begin{equation}
\hat\delta(\x,t) \simeq \sum_{i: \langle n_i\rangle
\gg1}\sqrt{2\langle n_i\rangle} \hat x_i(0)\psi_i(\x,t).
\end{equation}
Within this approximation any product of operators $\hat\delta$
depends only on $\hat x_i$ operators (in general it could also
depend on $\hat p_i=(\hat a^{ }_i-\hat a_i^\dagger)/i\sqrt2$) and
the $n$-th order correlation function,
\begin{eqnarray}\label{quant}
&\rho_n(\x_1,\ldots,\x_n;t) =  \langle\hat
\delta^\dagger(\x_1,t)\ldots\hat\delta^\dagger(\x_n,t)\hat\delta(\x_n,t)
\ldots\hat\delta(\x_1,t)\rangle\nonumber \\
&= \langle\{\hat
\delta^\dagger(\x_1,t)\ldots\hat\delta^\dagger(\x_n,t)\hat\delta(\x_n,t)
\ldots\hat\delta(\x_1,t)\}_{\mathrm{sym}}\rangle,
\end{eqnarray}
where the symmetrization is performed with respect to $\hat x_i$ and
$\hat p_i$. Thus the evaluation of the quantum average in
Eq.~(\ref{quant}) is equivalent to the procedure in which we replace
the operators $\hat x_i$'s and $\hat p_i$'s with c-number variables
$x_i$ and $p_i$ and perform an integral of the product of
$\delta(\x,t)=\sum_i\sqrt{2\langle n_i\rangle} x_i\psi_i(\x,t)$
function with the Wigner distribution of the vacuum state,
\begin{eqnarray}
&&\rho_n(\x_1,\ldots,\x_n;t)=\int \mathrm{d}\{x_i\}\int
\mathrm{d}\{p_i\}W(\{x_i\},\{p_i\})
\nonumber\\
&&\delta^*(\x_1,t)\ldots\delta^*(\x_n,t)\delta(\x_n,t)\ldots\delta(\x_1,t)=\label{Wigner}\\
&&\int \mathrm{d}\{x_i\}\int
\mathrm{d}\{p_i\}W(\{x_i\},\{p_i\})|\delta(\x_1,t)|^2
\ldots|\delta(\x_n,t)|^2.\nonumber
\end{eqnarray}
Here
\begin{equation}
W(\{x_i\},\{p_i\})=\prod_i
\frac{1}{\pi}\exp\left(-x_i^2-p_i^2\right).
\end{equation}
The integrals over $p_i$'s can be calculated explicitly and we
obtain
\begin{eqnarray}
&&\rho_n(\x_1,\ldots,\x_n; t)=\int \mathrm{d}\{x_i\}
\prod_i\frac{1}{\sqrt\pi}\exp\left(-x_i^2\right)\times\nonumber\\
&&|\delta(\x_1,t)|^2\cdot\ldots\cdot|\delta(\x_n,t)|^2
\end{eqnarray}
We stress that any correlation function can be evaluated according
to the procedure described above and conclude that
\begin{equation} \label{distrib}
P(\{x_i\})=\prod_i\frac{1}{\sqrt\pi}\exp\left(-x_i^2\right)
\end{equation}
gives the probability distribution of quadratures' amplitudes
$x_i$'s.

Thereby we generated a scheme for calculating the single realization
of the scattering experiment: we randomly choose the set of values
$x_i$ with probabilities $P(\{x_i\})$ and construct the
$\delta(\x,t)$ function. To find density distribution in a single
realization we take  $|\delta(\x,t)|^2$. This result is analogous to
that obtained in~\cite{Sacha}. Here we see the analogy to the
so-called Wigner stochastic method \cite{Norie}. In our method we
randomly choose the amplitudes of the quadratures of the
macroscopically populated modes and to calculate any moment of the
field operator $\hat\delta(\x,t)$ we simply perform a statistical
average over variables $x_i$. In contrary to the Wigner method, we
do not introduce any external noise to the dynamics of the system.
The detailed discussion will be given elsewhere.

Now we apply the method derived above to plot the density of the
hallo of scattered particles, in a single realization of the
collision. Since the density of particles is measured after the free
expansion of the atomic cloud, it represents the momentum
distribution of atoms. Hence we show the results in the momentum
space, $|\delta(\K,t)|^2$, where $\K$ is a wave-vector of the
scattered atoms.  We present both the cut as well as the column
density, i.e. $\int dk_z|\delta(\K,t)|^2$, see Fig.~\ref{fig1}. All
the data corresponds to the time $t$, long after the collision was
completed.

As shown in~\cite{Zin2} the angular size of the coherence of
scattered atoms is proportional to $1/\beta$. We expect that by
decreasing $\beta$, and thus increasing the range of coherence, at
some stage we should observe clear anisotropy (spikes) in the
distribution of scattered atoms. The spikes would be an effect of
the bosonic stimulation of scattering into highly occupied modes. On
the other hand, as $\beta$ increases, the range of coherence
decreases. Then, scattering of separate atoms is practically
independent. In such a case we expect that all the momentum modes
available to scattered atoms should be almost uniformly occupied.
These predictions agree very well with the results obtained in the
numerical simulations.

Figure \ref{fig1} shows the cut for $k_z=0$ as well as the column
density of atoms scattered out from the condensate for three
different values of $\beta$. The ratio $\alpha/\beta$, which is a
measure of bosonic enhancement~\cite{Zin2}, is in all cases the
same, and equals $\alpha/\beta=5$. This value corresponds to the
regime of strong bosonic enhancement, where condition of high mode
occupation is fulfilled. For the fixed radius of the condensate one
can vary the value of $\beta$ by changing the mutual velocity of the
colliding clouds. In Fig.~\ref{fig1}, case a), the collision can be
regarded as slow and we observe a clear speckle structure. For
larger values of $\beta$ (Figure \ref{fig1}b and \ref{fig1}c) the
distribution becomes more and more uniform and resembles an average
over many realizations.

Notice that the increase of $\beta$ can be achieved also by changing
the size of the clouds, which is proportional to number of atoms.
That explains the experimental results obtained in \cite{Vogels}.
They are consistent with our data. While their condensates where
rather slow the size ($3\times10^{7}$ atoms) was very big, giving
$\beta\simeq100$. In this regime we expect that after coarse
graining introduced by the CCD camera resolution, the momentum
density of scattered atoms would be highly uniform. This is exactly
what we see in experimental data shown in Fig.2b of \cite{Vogels}.
If one would intend to see the speckles, he should reduce the
velocity of colliding condensates and/or reduce their size by
decreasing the number of atoms. Alternatively one can use tighter
traps. According to our estimates it should be feasible to reduce
the value of $\beta$ to reach the $\beta\simeq20$ region.

Due to the large number of atoms, even the data obtained in a single
experiment has some attributes of the statistical average
~\cite{Bach}. It is illustrated in the last column of
Fig.\ref{fig1}. In this case we fixed the angle between
$\K_1=[\phi',\theta',Q]$ and $\K_2=[\phi'+\phi,\theta',Q]$, keeping
their length equal to $Q$ and calculated the average $\int
\delta^*(\K_1) \delta(\K_2)\,d\Omega'$ for three different
realizations (dashed lines) and compared it with quantum mechanical
average $\langle\delta^\dagger(\K_1) \delta(\K_2)\rangle$ (solid
line). The agreement is striking and getting better with increasing
value of $\beta$ (when the distribution becomes more uniform).

In summary we show that single realization of the experiment of two
colliding condensates contains an interesting information and may
differ substantially from the picture obtained after averaging over
many realizations. Especially for slow collisions we predict the
lumpy structure in the momentum distribution of scattered atoms,
with clear spikes even in the column density. The spikes found in
this Letter are analogous to the speckles in the light beam produced
by a multimode laser \cite{Laser}. Even closer optical analogy can
be found in recent studies of the parametric down conversion
\cite{Wasil}.

The authors acknowledge support from KBN Grant 2P03B4325,
1P03B14729, 1P03B14629, and the Polish Ministry of Scientific
Research and Information Technology under Grant No.
PBZ-MIN-008/P03/2003 (M.T.).

\begin{widetext}
\phantom{}
\begin{figure}[htb]
\centering
\includegraphics[scale=0.78, angle=0]{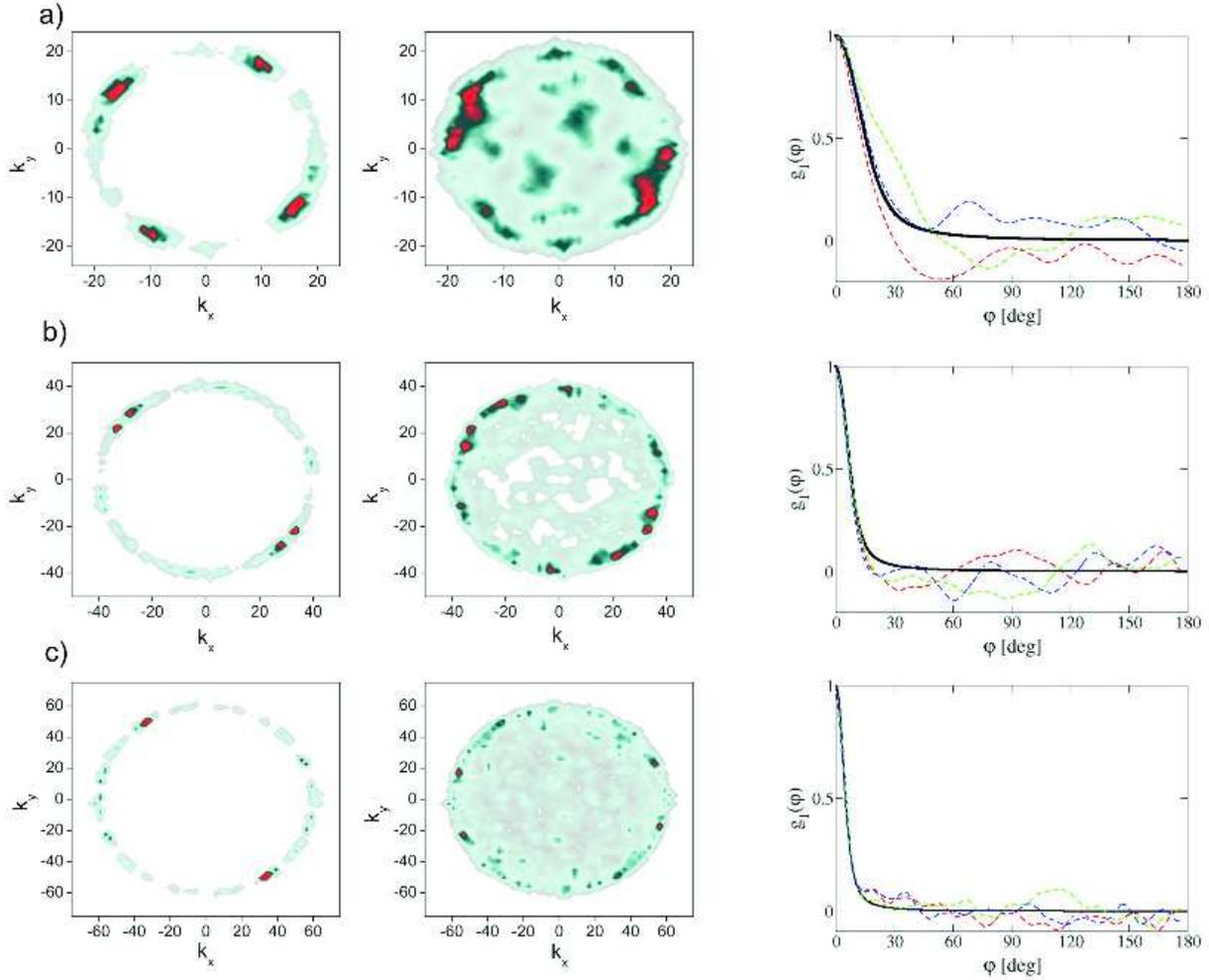}
\caption{(Color online): Momentum distribution and normalized
density matrix of scattered atoms for three different sets of
parameters $(\beta,\alpha)$; a) (20,100), b) (40,200), c) (60,300).
Left panel corresponds to a cut with $k_z=0$, center to a column
density. The right panel shows the comparison of the normalized
density matrix averaged over many realizations (solid line) with the
three different cases (dashed lines) of the density matrix
calculated for a single realization averaged over the full solid
angle.}\label{fig1}
\end{figure}
\end{widetext}


\begin{thebibliography}{20}
\bibitem{remark1} We work in Heisenberg picture, so all averages
are calculated in the vacuum state $|\Omega\rangle$:
$a_i|\Omega\rangle=0$ for all $i$'s.
\end{thebibliography}
\end{document}